\def\monthyear{\ifcase\month\or
  January\or February\or March\or April\or May\or June\or
  July\or August\or September\or October\or November\or
December\fi
  \space\number\year}

\renewcommand{\baselinestretch}{1.17}
\def\section{\@startsection {section}{1}{\z@}{-3.5ex plus -1ex
minus
 -.2ex}{2.3ex plus .2ex}{\large\bf}}
\def\subsection{\@startsection{subsection}{2}{\z@}{-3.25ex plus -
1ex minus
 -.2ex}{1.5ex plus .2ex}{\normalsize\bf}}

\newcommand{\gapproxeq}{\lower
.7ex\hbox{$\;\stackrel{\textstyle >}{\sim}\;$}}
\newcommand{\lapproxeq}{\lower
.7ex\hbox{$\;\stackrel{\textstyle <}{\sim}\;$}}

\newcounter{appendice}

\def\thefiglist#1{\section*{Figure Captions\markboth
 {FIGURE CAPTIONS}{FIGURE CAPTIONS}}\list
 {Figure \arabic{enumi}.}
 {\settowidth\labelwidth{Figure #1.}\leftmargin\labelwidth
 \advance\leftmargin\labelsep
 \usecounter{enumi}}
 \def\baselinestretch{1.1}\@normalsize
 \def\newblock{\hskip .11em plus .33em minus -.07em}
 \sloppy}

\documentstyle[12pt]{article}
\newcommand{\half}{\mbox{\small $\frac{1}{2}$}}

\newcommand{\Tr}{\,{\rm Tr}}

\newcommand{\hodge}{\mbox{\mbox{}$^*\!$}}
\begin{document}
\begin{titlepage}
\begin{flushright} RAL-93-079\\
hep-th/9312071
\end{flushright}
\vskip 1.5cm
\begin{center}
{\bf\Large Classical and Quantum Mechanics of Dirac-like Topological Charges
in Yang-Mills Fields}
\vskip 1cm
{\large Chan Hong-Mo}\\
\vskip .5cm
{\it Rutherford Appleton Laboratory,\\
Chilton, Didcot, Oxon, OX11 0QX, UK.}\\
\vskip .7cm
{\large J. Faridani}\\
\vskip .5cm
{\it Department of Theoretical Physics, Oxford University,\\
1 Keble Road, Oxford, OX1 3NP, UK.}\\
\vskip .7cm
{\large Tsou Sheung Tsun}\\
\vskip .5cm
{\it Mathematical Institute, Oxford University,\\
24-29 St. Giles', Oxford, OX1 3LB, UK.}\\
\end{center}
\begin{abstract}
Most nonabelian gauge theories admit the existence of conserved and quantized
topological charges as generalizations of the Dirac monopole.  Their
interactions are dictated by topology.  In this paper, previous work in
deriving classical equations of motion for these charges is extended to
quantized particles described by Dirac wave functions.  The resulting
equations show intriguing similarities to the Yang-Mills theory.  Further,
although the system is not dual symmetric, its gauge symmetry is nevertheless
doubled as in the abelian case from $G$ to $G \times G$, where the second $G$
has opposite parity to the first but is mediated instead by an antisymmetric
second-rank tensor potential.
\end{abstract}
\end{titlepage}

\baselineskip 24pt
\section{Introduction}

It is well-known that Yang-Mills theories, by virtue only of their intrinsic
gauge structure, admit the existence of a class of topological charges
which are automatically quantized and conserved.  In the special case of
an abelian theory, they are the Dirac monopoles \cite{Dirac,Lubkin,Wuyang1,
Coleman}.  Although in the abelian case the interactions of monopoles are
just the dual to those of ordinary charges, their generalizations to
nonabelian theories, for lack of a dual symmetry there, are still largely
unknown.  The object of this paper is to investigate in nonabelian gauge
theories the dynamical properties of particles carrying these charges with
the view of ascertaining eventually whether such particles might exist in
nature.

These charges appear as topological obstructions associated with nontrivial
$G$-bundles (for a theory with gauge or structure group $G$) over $S^2$.  One
way to realize this specifically is as follows \cite{Chantsou}.  Consider
the phase factor:
\begin{equation}
\Phi(C) = P \exp ig \oint_C A_\mu dx^\mu,
\label{PhiofC}
\end{equation}
where $A_\mu(x)$ is the Lie algebra-valued gauge potential, the integral
is taken over some closed loop $C$ in 4-dimensional space-time, and
the symbol $P$ denotes path-ordering, from right to left in our convention.
This $\Phi(C)$ maps each loop in space-time to an element of the gauge group
$G$.  Take now a 1-parameter family of closed loops $C_t$ labelled by a
parameter $t$ such that for $t = 0$ and $2\pi$ $C_t$ shrinks to a point
and as $t$ varies from $0$ to $2\pi$ $C_t$ envelops a closed 2-dimensional
surface $\Sigma$ in space-time as depicted in Figure \ref{loopover}. Then, for
\begin{figure}
\vspace{10cm}
\caption{Loops enveloping a closed surface}
\label{loopover}
\end{figure}
every value of $t$ we have a group element $\Phi(C_t)$ so that as $t$ varies
from $0$ to $2\pi$, $\Phi(C_t)$ traces out a closed curve, say $\Gamma$, in the
gauge group $G$, starting and ending at the group identity.  Suppose now
we continuously deform the surface $\Sigma$ to a point, what will happen
to the curve $\Gamma$ in $G$?  The answer will depend on the topology of
$G$.  If $G$ is simply connected, then the curve $\Gamma$ will also shrink
to a point.  However, if the gauge group $G$ is not simply connected, then
some curves $\Gamma$ cannot shrink to a point, in which case we say that there
is a topological obstruction inside $\Sigma$.  Such an obstruction cannot
be destroyed by any continuous variations of the field variables $A_\mu(x)$,
and can thus be pictured as a conserved charge.  The values this charge can
take are labelled by the (homotopy) classes of curves $\Gamma$ in $G$ which
cannot be continuously deformed into one another, and these are by definition
discrete or `quantized' as claimed.  In mathematical language, they are
the elements of the fundamental group $\pi_1(G)$, while gauge field
configurations labelled by nonzero elements of $\pi_1(G)$ correspond to
nontrivial $G$-bundles over the 2-dimensional surface $\Sigma$.

In the particular case of an abelian theory, the gauge group $U(1)$
has an infinite number of homotopy classes of closed curves labelled
by their winding numbers $n$ in $U(1)$,  i.e. $\pi_1(G) = {\bf Z}$.  It is
then readily seen that the topological obstructions described above
represent magnetic monopoles, with magnetic charges ${\tilde e} = n/2e$.
For this reason, it has been suggested that the analogous charges for
nonabelian
Yang-Mills fields be called ``nonabelian monopoles'', since they are the direct
generalizations of the same topological concepts.  However, this name has
also been applied to the 't\ Hooft-Polyakov solitons in Yang-Mills-Higgs
theories, which are actually abelian monopoles embedded in nonabelian
theories \cite{tHooft,Polyakov2}.  (In other words, these latter are
nontrivial $U(1)$-subbundles of (usually) trivial nonabelian bundles, while
those we are considering are nontrivial nonabelian bundles.)  To avoid
confusion, therefore, we shall refer instead to the topological charges
defined above as $\zeta$-charges.  This nomenclature serves also to
distinguish them from the ordinary `nontopological' charges such as `colour'
in QCD occuring as nonvanishing divergence $D^\nu F_{\mu\nu}$ to the gauge
field which will be referred to as `source charges' in what follows.

The values that $\zeta$-charges can take depend on the gauge group.
For pure Yang-Mills theories with gauge algebra ${\bf su(N)}$, (whose gauge
groups are $SU(N)/{\bf Z_N}$), $\zeta$ has the following range of values
\cite{Chantsou}:
\begin{equation}
\zeta_r = \exp (i2\pi r/N), \ \ \ r = 0, 1, ..., N-1.
\label{zetar}
\end{equation}
In particular, for the simple case of ${\bf su}(2)$ pure Yang-Mills theory
(with
gauge group $SU(2)/{\bf Z_2} = SO(3)$) to which we shall often refer, the
$\zeta$-charge can take only two values which may be denoted by a sign $\pm$,
where $+$ corresponds to the vacuum value.

As stated above, our aim here is to formulate and study the dynamics of
$\zeta$-charges with the view of eventually ascertaining whether such charges
may exist in Nature.  We think that such a question is meaningful since,
electrodynamics being `dual symmetric', an electric charge may be considered
either as a source charge in the Maxwell field $F_{\mu\nu}(x)$ as is usually
done, or else as a monopole or $\zeta$-charge in the dual field
$\hodge F_{\mu\nu}(x)$, which is also a gauge field.  Hence, $\zeta$-charges
of nonabelian fields are in a sense an equally valid generalization of the
electric charge as the `colour' source charges in usual Yang-Mills theory,
and may thus also lead to interesting physics.

The beauty of the problem is that the topological definition of $\zeta$-charges
already implies their interactions \cite{Wuyang2,Chanstsou2}.  That this is
the case can be seen intuitively as follows.  The presence of a $\zeta$-charge
at some point in space means that the gauge field in the spatial region
surrounding that point has a certain topological configuration.  Thus, if the
position of the $\zeta$-charge changes, while its value, being conserved,
remains unchanged, it follows that the gauge field will have to rearrange
itself so as to maintain a similar topological configuration about the new
point.  In physical language therefore, there is already an induced coupling
between the $\zeta$-charge's position and the field variables.

To discover the explicit form of this intrinsic interaction, one can proceed
as follows.  One begins by writing down the free action of the gauge field
and that of a particle carrying the $\zeta$-charge, thus:
\begin{equation}
{\cal A}^0 = {\cal A}^0_F + {\cal A}^0_M,
\label{freeaction}
\end{equation}
where ${\cal A}^0_F$ for the free field is usually taken to be:
\begin{equation}
{\cal A}^0_F = -\frac{1}{16\pi} \int d^4x \Tr[F_{\mu\nu}(x) F^{\mu\nu}(x)],
\label{freefielda}
\end{equation}
in which $F_{\mu\nu}(x)$ is the field tensor:
\begin{equation}
F_{\mu\nu}(x) = \partial_\nu A_\mu(x) - \partial_\mu A_\nu(x)
   + ig [A_\mu(x), A_\nu(x)].
\label{Fmunu}
\end{equation}
If one is dealing with a classical point particle with mass $m$, then one
would write for ${\cal A}^0_M$:
\begin{equation}
{\cal A}^0_M = - m \int d \tau \sqrt{\frac{dY^\mu(\tau)}{d \tau}
   \frac{dY_\mu(\tau)}{d \tau}},
\label{freecparta}
\end{equation}
where $Y^\mu(\tau)$ represents the world-line of the particle.  Extremizing
the action ${\cal A}^0$ with respect to the dynamical variables $A_\mu(x)$
and $Y^\mu(\tau)$ leads to the equations of motion of the free field and
the free particle.  Suppose however we now stipulate that the particle
at $Y^\mu(\tau)$ should carry a $\zeta$-charge.  In that case the gauge
potential $A_\mu(x)$ is required to be in a particular topological
configuration, the specification of which depends on the position
$Y^\mu(\tau)$ of the $\zeta$-charge.  If we now extremize ${\cal A}^0$ in
(\ref{freeaction}) with respect to $A_\mu(x)$ and $Y^\mu(\tau)$ under this
constraint, the equations of motion will contain `interactions', coupling
the particle co-ordinates $Y^\mu(\tau)$ to the field variables $A_\mu(x)$.

The above criterion for deriving the intrinsic interaction between a gauge
field and a $\zeta$-charge was first applied by Wu and Yang in 1976
\cite{Wuyang2} to the abelian theory and a classical point particle.
The result was the standard Maxwell and Lorentz equations for a Dirac
monopole interacting with the electromagnetic field, exactly as
expected from dual symmetry.  The same criterion can in principle
be applied to any Yang-Mills theory and to any physical entity carrying
the $\zeta$-charge to derive their interactions \cite{Chanstsou2}.  However,
for nonabelian theories, the result will be new, since for lack of a dual
symmetry, the interactions of $\zeta$-charges there can no longer be
surmised from those of source charges.

Using the above criterion, we derive in this paper the equations of motion
for both classical and quantum point particles carrying $\zeta$-charges in
Yang-Mills fields.  The actual objective is the quantum equations for the
nonabelian case since the abelian equations are known by dual symmetry while
the classical nonabelian problem has already been solved in an earlier paper
\cite{Chanstsou2}.  However, in order to derive the quantum equations for
the nonabelian case, we found it necessary both first to examine how the
quantum equations can be derived in the abelian case using the present
criterion, and then to reformulate and re-solve the classical nonabelian
problem by a new approach which could be adapted to the quantum theory.  To
avoid technical difficuties due to patching of gauge potentials, we use a loop
space formalism developed earlier for the purpose, the result of which
we shall briefly summarize.  With the view of making the intriguing comparison
with the standard Yang-Mills theory for source charges, the quantum equations
for $\zeta$-charges are worked out here explicitly for a Dirac particle,
although it is in principle possible to do the same for other quantum
particles.

The equations so obtained display some interesting novel features which
are all generalizations, but in a quite nontrivial fashion, of familiar
features in the abelian case.

(a) The abelian theory is dual symmetric, so that both source (electric) and
monopole (magnetic) charges take values which are integral multiples
of some unit charge; in other words, both sources and monopoles form
representations of the gauge group $U(1)$.  In the nonabelian theory,
there is no similar dual symmetry, and source and $\zeta$-charges are
defined very differently, with sources forming representations of the gauge
group $G$ (e.g. quarks are colour triplets in QCD), while $\zeta$-charges
are labelled just by elements (homotopy classes) of the fundamental group
$\pi_1(G)$.  However, it turns out that when the above procedure is
implemented to derive equations of motion, the $\zeta$-charge has to be
assigned an orientation in internal symmetry space, partly restoring the
similarity  with source charges.  Thus, a classical point particle
carrying a $\zeta$-charge would acquire a `dynamical charge' labelled
by an element of the gauge algebra, just like a source charge in Wong's
classical limit of the Yang-Mills theory \cite{Wong}, while a quantum particle
carrying a $\zeta$-charge would be described by a wave function belonging
to some representation of the gauge group $G$, again analogous to a
source charge in the standard Yang-Mills case.

(b) In the dual symmetric abelian theory, the dynamics is invariant
under arbitrary local rotations of both the phases of source (electric)
and monopole (magnetic) charge, so that the actual symmetry is enlarged
to $U(1) \times U(1)$.  In the nonabelian theory, no similar dual symmetry
applies.  Nevertheless, it is found that the dynamics has still this
enlarged $SU(N) \times SU(N)$ gauge symmetry, with the new $SU(N)$ symmetry
having, as in the abelian theory, the opposite parity to the original $SU(N)$.
This new symmetry is traced to the fact that the $\zeta$-charge was originally
defined independently of orientation in internal space.  Hence, the dynamics
ought in the end also to be invariant under arbitrary local redefinitions of
this orientation or `phase'.

(c) In the abelian theory, the dual field $\hodge F_{\mu\nu}(x) =
-(1/2) \epsilon_{\mu\nu\rho\sigma} F^{\rho\sigma}$ is also a gauge
field derivable from a potential, thus: $\hodge F_{\mu\nu}(x) = \partial_\nu
{\tilde A}_\mu(x) - \partial_\mu {\tilde A}_\nu(x)$, where ${\tilde A}_\mu(x)$
acts as parallel transport for the phase of the monopole.  For a nonabelian
theory, $ \hodge F_{\mu\nu}(x)$ is usually not a gauge field derivable from a
potential.  Nevertheless, it turns out that there is another quantity
emerging as the Lagrange multiplier for the defining constraint of the
$\zeta$-charge which acts as the parallel `phase' transport for the
$\zeta$-charge.  In contrast to usual gauge potentials such as $A_\mu(x)$,
however, this new parallel transport carries two space-time indices, and
transforms as the tensor potential first discovered in string and supersymmetry
theory \cite{Hayashi,Kalbramond,Cremmerscherk,Nambu}, in particular that
suggested by Freedman and Townsend in 1981 \cite{Freetown}.

(d) The `dual' potential ${\tilde A}_\mu(x)$ in the abelian theory
couples, of course, to a monopole in the same way as the usual potential
$A_\mu(x)$ couples to a source.  In the nonabelian theory, it is found
that there is still a quantity ${\tilde A}_\mu(x)$ which couples to the
$\zeta$-charge in exactly the same way that the ordinary Yang-Mills
potential $A_\mu(x)$ couples to a source charge.  However, this
${\tilde A}_\mu(x)$ is not related to $\hodge F_{\mu\nu}(x)$ as $A_\mu(x)$ is
to $F_{\mu\nu}(x)$, but is obtained instead as a directional average of
the tensor potential referred to in (c) above.

One sees therefore that although there is no dual symmetry in nonabelian
Yang-Mills theory, $\zeta$-charges still seem to function as a sort of
mirror image of ordinary source charges, though with somewhat different
dynamics.  It could thus be interesting to inquire how particles carrying
such charges will behave if they exist and how different they will be
compared to particles carrying ordinary source charges.  The question can
in principle be answered by studying the equations derived in this paper.
We have begun to do so but have as yet little to report.

\section{The Abelian Theory}

To learn how equations of motion are to be derived for quantum particles
carrying $\zeta$-charges in general, we need first to examine the special
case of the abelian theory where we know at least what equations to expect.

Let us first recall the previous solution for a classical point particle.
As outlined above, the equations of motion of a $\zeta$-charge are to be
determined by extremizing the free action (\ref{freeaction}), for
${\cal A}^0_F$ as given in (\ref{freefielda}) and ${\cal A}^0_M$ as
given in (\ref{freecparta}), with respect to the variables $A_\mu(x)$
and $Y^\mu(\tau)$, but under the topological constraint stipulating that
the particle at $Y^\mu(\tau)$ carries a specified $\zeta$-charge.  This
problem was first posed and solved, though not in the following manner,
by Wu and Yang in 1976 \cite{Wuyang2}.

As it stands, the problem is technically a little complicated, for two
reasons.  First, the topological constraint defining the $\zeta$-charge
is so far only abstractly given in terms of homotopy classes of closed
curves in the gauge group and is thus a little awkward to implement.  Second,
in the presence of a $\zeta$-charge, which is in this case the familiar
monopole, we know that the gauge potential $A_\mu(x)$ must have singularities
(the `Dirac string') \cite{Dirac}, or else in the language of Wu and Yang
\cite{Wuyang1}, it has to be patched, with patches dependent on the position
of the charge; this makes, in either language, the variational problem
directly in terms of $A_\mu(x)$ rather complicated.

To surmount these difficulties, a tactic was suggested \cite{Chanstsou2} which
we shall extend to other cases.  In general terms, this tactic consists of two
steps: first to reduce the topological constraint on (\ref{freeaction})
defining the $\zeta$-charge into an explicit local form, and second, to
replace the gauge potential $A_\mu(x)$ as field variable by patch-independent
quantities.  For the simple example of a classical $\zeta$-charge in an
abelian theory, these steps were implemented as follows.

First, we note that for the abelian theory, the path-ordering denoted by $P$ in
(\ref{PhiofC}) can be ignored so that $\Phi(C)$ is indeed an exponential
of a line integral which, by means of the Stokes' theorem can be written as
the magnetic flux passing through any surface bounded by $C$.  Hence, the
winding number denoting the homotopy class which represents the $\zeta$-charge
enveloped by the surface in Figure \ref{loopover} is just $4\pi$ times the
total magnetic flux emerging from that surface, or in other words the
magnetic charge as anticipated.  Since this is true for any surface
surrounding our point $\zeta$-charge or monopole, it follows from Gauss'
theorem that the topological constraint we wish to impose on the action
(\ref{freeaction}) can be written as the local condition:
\begin{equation}
\partial_\nu \hodge F^{\mu\nu}(x) = -4\pi {\tilde e} \int d \tau
   \frac{dY^\mu(\tau)}{d\tau} \delta^4(x-Y(\tau)),
\label{gausslawc}
\end{equation}
which is to be satisfied for all $x$ and $\mu$.

Second, we note that we can in this case replace as field variables the gauge
potential $A_\mu(x)$ by the field tensor $F_{\mu\nu}(x)$.  Normally, this is
not possible, since an arbitrary choice of an antisymmetric tensor
$F_{\mu\nu}(x)$ will not be derivable from a potential as it should.  However,
if we require $F_{\mu\nu}(x)$ to satisfy the constraint (\ref{gausslawc}) as
we wish to here, then except on the monopole world-line, the Bianchi identity
is satisfied, so that by the Poincar\'e lemma, $F_{\mu\nu}(x)$ is guaranteed
to be derivable from a potential.  And $F_{\mu\nu}(x)$, being gauge invariant,
is independent of patching as desired.

With these two observations, the solution of the variational problem posed
above becomes very simple.  Incorporating the constraint (\ref{gausslawc})
by means of Lagrange multipliers $\lambda_\mu(x)$, we construct the auxiliary
action:
\begin{equation}
{\cal A} = {\cal A}^0 + \int d^4 x \lambda_\mu(x) \{\partial_\nu
\hodge F^{\mu\nu}(x)
   + 4\pi {\tilde e} \int d\tau \frac{dY^\mu(\tau)}{d\tau}
   \delta^4(x-Y(\tau))\}.
\label{actcpart}
\end{equation}
Extremizing then with respect to the variables $F_{\mu\nu}(x)$ and
$Y^\mu(\tau)$
we obtain respectively:
\begin{equation}
F^{\mu\nu}(x) = 4 \pi \{\half \epsilon^{\mu\nu\rho\sigma}
   (\partial_\sigma \lambda_\rho(x) - \partial_\rho \lambda_\sigma(x))\},
\label{eleq1}
\end{equation}
and:
\begin{equation}
m \frac{d^2Y_\mu(\tau)}{d\tau^2} = -4\pi {\tilde e}
   \{\partial_\nu \lambda_\mu(Y(\tau)) - \partial_\mu \lambda_\nu(Y(\tau))\}
   \frac{dY^\nu(\tau)}{d\tau}.
\label{eleq2}
\end{equation}
The first equation (\ref{eleq1}) says that $\hodge F_{\mu\nu}(x)$ is also
a gauge field derivable from a gauge potential:
\begin{equation}
{\tilde A}_\mu(x) = 4\pi \lambda(x),
\label{atilde}
\end{equation}
and allows the Lagrange multiplier $\lambda_\mu(x)$ to be eliminated from
the equations, giving for (\ref{eleq2}):
\begin{equation}
m \frac{d^2Y^\mu(\tau)}{d\tau^2} = - {\tilde e} \hodge F^{\mu\nu}(Y(\tau))
   \frac{dY_\nu(\tau)}{d\tau}.
\label{loreqstar}
\end{equation}
In other words, we obtain exactly the dual of the standard equations for
a classical point source (i.e. electric) charge, as asserted above.

Let us see now how the same considerations can be extended to the quantum case
\cite{Chantsou}.  Consider then a quantum point particle described by a
wave function $\psi(x)$ and carrying a $\zeta$-charge in an abelian gauge
field.  In anticipation of future applications and for comparison with
standard Yang-Mills theory, let us work with a Dirac particle described
by a Dirac wave function, although the same method can in principle be applied
to, for example, a Schr\"{o}dinger particle.  Then, instead of the action
${\cal A}^0_M$ of (\ref{freecparta}), we write:
\begin{equation}
{\cal A}^0_M = \int d^4x {\bar \psi}(x) (i\partial_\mu \gamma^\mu - m) \psi(x),
\label{freeqparta}
\end{equation}
and instead of the constraint (\ref{gausslawc}) defining the $\zeta$-charge of
the classical particle, we propose to write:
\begin{equation}
\partial_\nu \hodge F^{\mu\nu}(x) = - 4\pi {\tilde e} {\bar \psi}(x) \gamma^\mu
   \psi(x),
\label{gausslawq}
\end{equation}
where we have just replaced the previous classical current by its quantum
analogue.  As in the classical case, we shall still use $F_{\mu\nu}(x)$ as
variables instead of the gauge potential so as to avoid complications with
patching.

Again the constraint (\ref{gausslawq}) can be incorporated by means of
Lagrange multipliers into an auxiliary action:
\begin{equation}
{\cal A} = {\cal A}^0 + \int d^4x \lambda_\mu(x) \{\partial_\nu
\hodge F^{\mu\nu}(x)
   + 4\pi {\tilde e} {\bar \psi}(x) \gamma^\mu \psi(x) \}.
\label{actqpart}
\end{equation}
Extremizing with respect to $F_{\mu\nu}(x)$, we obtain (\ref{eleq1}) as before,
and then with respect to $\psi(x)$, the equation:
\begin{equation}
(i \partial_\mu \gamma^\mu - m) \psi(x) = -{\tilde e} {\tilde A}_\mu(x)
   \gamma^\mu \psi(x),
\label{eleq3}
\end{equation}
with ${\tilde A}_\mu(x)$ defined in (\ref{atilde}), as before.  However, in
contrast to the classical case, the Lagrange multiplier can no longer be
eliminated from the equations of motion but remains behind obligingly to serve
as a `dual potential' coupling to the wave function as a gauge potential
should.

One sees thus that the method works also for the quantum particle giving
equations which are exactly the duals of the corresponding standard equations
for a point source as one expects.  We note a couple of points which will
be relevant for the future.  First, the constraint (\ref{gausslawq}) may look a
little disturbing when viewed as a generalization of the topological
definition of the $\zeta$-charge, which ought to be quantized.  This means
if we take the $\mu = 0$ component of (\ref{gausslawq}), then an integral
of the right-hand side over any volume should always give the value
$4\pi {\tilde e}$ times an integer, which it clearly does not for a
general $\psi(x)$.  For this reason, we believe that for a truly consistent
interpretation of (\ref{gausslawq}), $\psi(x)$ ought to be considered
as a second quantized field, so that for $\mu = 0$, the right-hand side
of (\ref{gausslawq}) will be just a numerical constant times the number
operator counting the (integral) number of $\zeta$-charges occuring at
$x$, whose integral over any volume will then be quantized.  Attempts
at second quantizing the theory are underway; here, however, we
shall take (\ref{gausslawq}) only at face value.  Second, we notice that
the equations of motion we deduced for the quantum $\zeta$-charge in an
abelian field has the further gauge symmetry:
\begin{equation}
\psi(x) \longrightarrow \exp \{i{\tilde e} {\tilde \Lambda}(x)\} \  \psi(x),
\label{psitransft}
\end{equation}
\begin{equation}
{\tilde A}_\mu(x) \longrightarrow {\tilde A}_\mu(x) + \partial_\mu
   {\tilde \Lambda}(x),
\label{atransft}
\end{equation}
in addition to the original gauge symmetry associated with $A_\mu(x)$.
This arose as a degeneracy in the solution of the Euler-Lagrange problem via
the Lagrange multiplier $\lambda_\mu(x)$, in spite of the fact that we were
solving the variational problem entirely in terms of quantities $\psi(x)$
and $F_{\mu\nu}(x)$ which ($\psi$ being `electrically' neutral) were both
invariant under the original symmetry.  As a result, we have a doubling of the
symmetry from the original $U(1)$ to $U(1) \times U(1)$ where the second
$U(1)$ has parity opposite to that of the first $U(1)$ because of the
\hodge-operation in the relation between $F_{\mu\nu}(x)$ and its dual.

\section{Loop Space}

In generalizing the above considerations to nonabelian Yang-Mills theories,
the same two technical difficulties mentioned at the beginning of the last
section still stand.  However, the tactics used successfully for solving
the abelian problem are no longer valid without modification.  First,
because of the path ordering $P$ in (\ref{PhiofC}), the quantity $\Phi(C)$,
in spite of the notation, is not really the exponential of a line integral
for which one can apply Stokes' theorem.  Indeed, even the concept of `flux'
has no direct generalization in the nonabelian theory.  We cannot therefore
reduce as we did for the abelian theory the defining topological condition
of the $\zeta$-charge into the local form (\ref{gausslawc}) or
(\ref{gausslawq}).  Secondly, the field tensor $F_{\mu\nu}(x)$ being now
covariant, rather than invariant as in the abelian case, is no longer
patch-independent; it is not useful therefore as replacement for the gauge
potential $A_\mu(x)$ as field variables.

A modification which suggests itself is to work in loop space
\cite{Polyakov1,Chanstsou1,Chantsou}.  First, the definition of the
$\zeta$-charge was originally given in terms of the loop quantity $\Phi(C)$ of
(\ref{PhiofC}) in any case.  Secondly, loop quantities are by definition
patch-independent, having thus the virtue we sought as replacement
variables to $A_\mu(x)$.  However, the problem with loop space
treatment is the high degree of redundancy in the field variables, which
makes manipulations in it rather unwieldy in general.  But in the particular
problem which interests us here, it turns out that several happy `coincidences'
combine to make loop variables particularly suitable.  The necessary
techniques have already been developed in an earlier work \cite{Chanstsou1}.
Here, we shall just summarize the results we need.

We shall work with parametrized loops, i.e. maps of the circle to
(4-dimensional) space-time, represented by functions:
\begin{equation}
\xi^\mu(s), \ \ s= 0 \rightarrow 2\pi, \ \ \xi^\mu(0) = \xi^\mu(2\pi).
\label{xiofs}
\end{equation}
It is sufficient to work only with loops all passing through a fixed reference
point $P_0$, so that $\xi$ is restricted to $\xi(0) = \xi(2\pi) = P_0$
The phase factor $\Phi(C)$ of (\ref{PhiofC}) is then a functional of $\xi$:
\begin{equation}
\Phi[\xi] = P_s \exp ig\int_0^{2\pi} ds A_\mu(\xi(s)) \dot{\xi}^\mu(s),
\label{Phiofxi}
\end{equation}
where a dot denotes differentiation with respect to the loop parameter $s$.

Following Polyakov \cite{Polyakov1}, we define next the quantity:
\begin{equation}
F_\mu[\xi|s] = \frac{i}{g} \Phi[\xi]^{-1} \frac{\delta}{\delta \xi^\mu(s)}
   \Phi[\xi],
\label{Fmuxis}
\end{equation}
which may be expressed in terms of local field quantities as:
\begin{equation}
F_\mu[\xi|s] = \Phi_\xi^{-1}(s_-,0) F_{\mu\nu}(\xi(s_-)) \Phi_\xi(s_-,0)
   \dot{\xi}^\nu(s),
\label{Fmuxisx}
\end{equation}
where $\Phi_\xi(s_- ,0)$ is the parallel transport from the reference point
$P_0$ at $s=0$ to the point $\xi(s_-)$ along the loop $\xi$:
\begin{equation}
\Phi_\xi(s_2, s_1) = P \exp ig \int_{s_1}^{s_2} ds' A_\mu(\xi(s'))
   \dot{\xi}^\mu(s'),
\label{Phixis1s2}
\end{equation}
and the subscript in $s_-$ means that in case of ambiguity the value of
$\xi(s)$
has to be taken as the limit from below, namely $\xi(s_-) = \xi(s - \epsilon),
\epsilon \rightarrow 0+$.  Whence it is readily seen that $F_\mu[\xi|s]$
depends on $\xi(s')$ only for $s' \le s$, or equivalently:
\begin{equation}
\frac{\delta}{\delta \xi^\nu(s')} F_\mu[\xi|s] = 0,\ \ \ s' > s,
\label{Fuptos}
\end{equation}
which is a consequence of our ordering convention, and that it has only
components transverse to the loop, namely:
\begin{equation}
F_\mu[\xi|s] \dot{\xi}^\mu(s) = 0,
\label{transverseF}
\end{equation}
which follows from the antisymmetry of $F_{\mu\nu}(x)$, or equivalently from
the invariance of $\Phi[\xi]$ under reparametrization.  In what follows, both
these properties (\ref{Fuptos}) and (\ref{transverseF}) of $F_\mu[\xi|s]$,
which basically just restrict the range of its arguments and the number of
its components, will be regarded as understood and absorbed into the notation.

It was proposed to adopt these quantities $F_\mu[\xi|s]$ as replacements
for the gauge potentials $A_\mu(x)$ as field variables in analogy with the
field tensor $F_{\mu\nu}(x)$ for the abelian case.  As such, $F_\mu[\xi|s]$
has the following attractive features.  First, it is closely related to
$F_{\mu\nu}(x)$ by (\ref{Fmuxisx}).  Second, similar to $F_{\mu\nu}(x)$
in the abelian theory, it is gauge invariant (apart from an $x$-independent
rotation at the reference point $P_0$ which is easily handled and for
convenience of presentation will henceforth be ignored) and therefore
patch-independent.  Third, the $\zeta$-charge, which was so far defined
for nonabelian theories only abstractly as a homotopy class, can be
expressed explicitly in terms of $F_\mu[\xi|s]$, as follows.

Geometrically, $F_\mu[\xi|s]$ may be interpreted from its definition in
(\ref{Fmuxis}) as a connection in loop space prescribing parallel transport
of the phases $\Phi[\xi]$ from a loop to neighbouring loops.  It can thus be
used, in close analogy to (\ref{PhiofC}) for $A_\mu(x)$, to construct the
holonomy for a loop in loop space, which when viewed in ordinary space-time
is a 2-dimensional closed surface as that depicted in Figure \ref{loopover}
introduced before to define the $\zeta$-charge.  Indeed, if we take a
1-parameter family of parametrized loops:
\begin{equation}
\Sigma = \{\xi_t^\mu(s); s = 0 \rightarrow 2\pi, t = 0 \rightarrow 2\pi,
   \xi_t(0) = \xi_t(2\pi) = \xi_0(s) = \xi_{2\pi}(s) = P_0\}
\label{ximuts}
\end{equation}
satisfying the specified boundary conditions so that as $t$ varies
from $0$ to $2\pi$, the loops $\xi_t$ envelop the 2-dimensional surface
$\Sigma$, then it can be shown that the $\zeta$-charge enclosed inside
$\Sigma$ can be labelled by the loop space holonomy:
\begin{equation}
\Theta_\Sigma = \zeta_\Sigma,
\label{Gausslawl}
\end{equation}
where, analogously to (\ref{Phiofxi}):
\begin{equation}
\Theta_\Sigma = P_t \int_0^{2\pi} dt \int_0^{2\pi} ds F_\mu[\xi_t|s]
   \frac{\partial \xi_t^\mu(s)}{\partial t}.
\label{Thetasigma}
\end{equation}
Thus, for a pure ${\bf su}(N)$ theory with gauge group $SU(N)/{\bf Z}_N$,
$\Theta_\Sigma$ takes values in ${\bf Z}_N$, whose elements (\ref{zetar})
also label the homotopy classes of closed curves in $G$.

Like all loop variables, however, $F_\mu[\xi|s]$ forms a highly redundant
set, which has to be severely constrained.  And here occurs one of those
`coincidences' which makes loop space techniques particularly suitable for
the present problem \cite{Chanstsou1,Chanstsou2}.  Indeed, it so turns out
that there is an extension of the (abelian) Poincar\'e lemma to the general
case which says that, given a set of these variables $F_\mu[\xi|s]$, then
so long as they are required to satisfy the relation (\ref{Gausslawl}) above
for isolated $\zeta$-charges, they will be expressible in terms of some
$A_\mu(x)$ via (\ref{Phiofxi}) and (\ref{Fmuxis}), except of course at the
locations of the $\zeta$-charges.  This means that if we were to impose the
condition (\ref{Gausslawl}) as a dynamical constraint, then in close analogy
to (\ref{gausslawc}) in the abelian case, the redundancy inherent in
$F_\mu[\xi|s]$ as variables will also be removed.  But this is exactly
the sort of constraint we wish to impose to derive the equations of motion
according to the criterion promulgated above.

The assertion (\ref{Gausslawl}) can alternatively be stated in a loop space
local form using the curvature constructed with $F_\mu[\xi|s]$ as connection
\cite{Polyakov1}, namely:
\begin{equation}
G_{\mu\nu}[\xi|s] = \frac{\delta}{\delta \xi^\nu(s)} F_\mu[\xi|s] -
  \frac{\delta}{\delta \xi^\mu(s)} F_\nu[\xi|s] +
ig[F_\mu[\xi|s],F_\nu[\xi|s]].
\label{Gmunuxis}
\end{equation}
This represents the phase change over a little loop in loop space, or in
4-dimensional space-time over a little closed surface enveloping a 3-volume,
as illustrated in Figure \ref{smallloop}.
\begin{figure}
\vspace{10cm}
\caption{Illustration for $G_{\mu\nu}[\xi|s]$ in ordinary space}
\label{smallloop}
\end{figure}
In terms of ordinary space-time variables, $G_{\mu\nu}[\xi|s]$ takes the
following form \cite{Chanstsou1}:
\begin{equation}
G_{\mu\nu}[\xi|s] = \Phi_\xi^{-1}(s_-,0) \epsilon_{\mu\nu\rho\sigma}
   D_\alpha \hodge F^{\rho\alpha}(\xi(s)) \Phi_\xi(s_-,0) \dot{\xi}^\sigma(s),
\label{Gmunuxisx}
\end{equation}
where $D_\mu$ is the ordinary gauge covariant derivative and, as in
(\ref{Fmuxisx}), the subscript in $s_-$ again means that in
case of ambiguity the value of $\xi(s)$ has to be taken as the limit from
below, as indicated in Figure \ref{smallloop}.  From (\ref{Gmunuxisx}), one
sees that $G_{\mu\nu}[\xi|s]$ vanishes unless the surface in Figure
\ref{smallloop} encloses a $\zeta$-charge.  Its value when it hits a
$\zeta$-charge is an element of the gauge algebra $\kappa$, which according
to (\ref{Gausslawl}) above, satisfies the condition:
\begin{equation}
\exp (i\pi \kappa) = \zeta.
\label{kappa}
\end{equation}
It is this `local' form of the expression for the $\zeta$-charge that we
shall use below.

Having resolved the question of redundancy of loop variables, one can then
in principle reformulate Yang-Mills theory entirely in terms of them.  In
particular, it can be seen from (\ref{Fmuxisx}) that the free field action
(\ref{freefielda}) can be written as:
\begin{equation}
{\cal A}_F^0 = -\frac{1}{4 \pi {\bar N}} \int \delta \xi \int_0^{2\pi} ds
   \Tr \{F_\mu[\xi|s] F^\mu[\xi|s]\} \dot{\xi}^{-2}(s),
\label{freefieldal}
\end{equation}
with:
\begin{equation}
{\bar N} = \int_0^{2\pi} ds \int \prod_{s' \neq s} d^4 \xi(s').
\label{Nbar}
\end{equation}

\section{Classical Mechanics of $\zeta$-Charge}

Although the equations of motion for a classical particle carrying a
$\zeta$-charge have already been derived in an earlier work \cite{Chanstsou2},
we need to develop here a new approach for deriving the same equations which
we can then extend to the quantum case.

As suggested above, the dynamics is given by a constrained variational
principle in which the free action is extremized under the constraint
provided by the defining condition of the $\zeta$-charge.  For a classical
point
particle, the action is given as in (\ref{freeaction}), (\ref{freefielda})
and (\ref{freecparta}).  But, in view of the observations at the beginning of
the last section we shall work rather with the field action in its loop space
form, namely (\ref{freefieldal}) instead of (\ref{freefielda}).

The defining condition of the $\zeta$-charge can be written in loop space in
either the global form (\ref{Gausslawl}) or in a differential form in terms
of the curvature $G_{\mu\nu}[\xi|s]$, both of which can be used to drive the
equations of motion.  Previously, the equations were derived using the global
form; here we shall rederive the result by imposing instead the differential
form, which we then extend to the quantum case.  For a $\zeta$-charge moving
along the world-path $Y^\mu(\tau)$ then, we write the constraint as:
\begin{equation}
G_{\mu\nu}[\xi|s] = 4\pi {\tilde g} \kappa[\xi|s] \epsilon_{\mu\nu\rho\sigma}
   \dot{\xi}^\rho(s) \int d\tau \frac{dY^\sigma(\tau)}{d\tau}
   \delta^4(\xi(s)-Y(\tau)).
\label{Gausslawldc}
\end{equation}
If one substitutes (\ref{Gmunuxisx}) into (\ref{Gausslawldc}), one obtains
in parallel to (\ref{gausslawc}) for the abelian case:
\begin{equation}
D_\nu \hodge F^{\mu\nu}(x) = - 4 \pi {\tilde g} \int d\tau
   K(\tau) \frac{d Y^\mu(\tau)}{d\tau} \delta^4(x - Y(\tau)),
\label{Gausslawdc}
\end{equation}
with:
\begin{equation}
K(\tau) = \left. \Phi_\xi(s_-,0) \kappa[\xi|s] \Phi_\xi^{-1}(s_-, 0)
   \right |_{\xi(s) = Y(\tau)},
\label{Ktau}
\end{equation}
which is in appearance similar to the Wong equation \cite{Wong} for a
classical Yang-Mills point source, although its content is different.  We note
in particular that in analogy to a source the particle at $Y(\tau)$ has now
acquired an orientation in internal symmetry space through the algebra
element $K(\tau)$ which a $\zeta$-charge did not originally possess.

Incorporating the constraint (\ref{Gausslawldc}) into the action by means
of Lagrange multipliers $L_{\mu\nu}[\xi|s]$, we have:
\begin{equation}
{\cal A} = {\cal A}^0 + \int \delta \xi ds \Tr \{L^{\mu\nu}[\xi|s]
   (G_{\mu\nu}[\xi|s] + 4\pi J_{\mu\nu}[\xi|s])\},
\label{Fullact}
\end{equation}
where $J_{\mu\nu}[\xi|s]$ is $-1/4\pi$ times the right-hand side of
(\ref{Gausslawldc}).
Equations of motion are now to be obtained by extremizing (\ref{Fullact})
with respect to free variations of $F_\mu[\xi|s]$ and $Y^\mu(\tau)$.  Varying
with respect to $F_\mu[\xi|s]$, one has:
\begin{equation}
(4\pi {\bar N} \dot{\xi}^2(s))^{-1} F_\mu[\xi|s]
   = - {\cal D}^\nu(s) L_{\mu\nu}[\xi|s],
\label{ELeq1}
\end{equation}
where
\begin{equation}
{\cal D}_\nu(s) = \frac{\delta}{\delta \xi^\nu(s)} - ig [F_\nu[\xi|s],\ \ \ ]
\label{covderivl}
\end{equation}
denotes the covariant derivative in loop space with $F_\mu[\xi|s]$ as
connection, and with respect to $Y^\mu(\tau)$:
\begin{equation}
m \frac{d^2Y_\mu(\tau)}{d\tau^2} = - 8\pi{\tilde g} \int \delta \xi ds
   \epsilon_{\mu\nu\rho\sigma} \frac{\delta}{\delta \xi^\lambda(s)}
   \Tr\{L^{\lambda\rho}[\xi|s] \kappa[\xi|s]\} \frac{dY^\nu(\tau)}{d\tau}
   \dot{\xi}^\sigma(s) \delta^4(\xi(s)-Y(\tau)).
\label{ELeq2}
\end{equation}
Together with the original constraint equation (\ref{Gausslawldc}),
(\ref{ELeq1}) and (\ref{ELeq2}) represent the equations of motion in
parametric form depending on the Lagrange multiplier $L_{\mu\nu}[\xi|s]$.

These equations can be rearranged to take more familiar forms.  First,
differentiating (\ref{ELeq1}) with respect to $\xi^\mu(s)$ we have:
\begin{equation}
(2 \pi {\bar N} \dot{\xi}^2(s))^{-1} {\cal D}^\mu(s) F_\mu[\xi|s]
   = - [{\cal D}^\mu(s), {\cal D}^\nu(s)] L_{\mu\nu}[\xi|s].
\label{propolyakov}
\end{equation}
Hence, using on the right-hand side the familiar relation between
the covariant derivative and the curvature, we deduce that:
\begin{equation}
(2 \pi {\bar N} \dot{\xi}^2(s))^{-1} \frac{\delta}{\delta \xi_\mu(s)}
   F_\mu[\xi|s] = -ig [G^{\mu\nu}[\xi|s], L_{\mu\nu}[\xi|s]],
\label{Expolyakov}
\end{equation}
Except at the position of the $\zeta$-charge, this implies by
(\ref{Gausslawldc}) the Polyakov equation \cite{Polyakov1}:
\begin{equation}
\frac{\delta}{\delta \xi_\mu(s)} F_\mu[\xi|s] = 0,
\label{Polyakov}
\end{equation}
which by (\ref{Fmuxisx}) is equivalent to the Yang-Mills equation:
\begin{equation}
D^\nu F_{\mu\nu}(x) = 0.
\label{Yangmills}
\end{equation}
Secondly, we note that in (\ref{ELeq2}), we may write:
\begin{equation}
\frac{\delta}{\delta \xi^\lambda(s)} \Tr\{L^{\lambda\rho}[\xi|s]
\kappa[\xi|s]\}
   = \Tr\{({\cal D}_\lambda(s) L^{\lambda\rho}[\xi|s])\kappa[\xi|s]\}
   + \Tr\{L^{\lambda\rho}[\xi|s]({\cal D}_\lambda(s) \kappa[\xi|s])\}.
\label{dtracelk}
\end{equation}
However, the Bianchi identity for $G_{\mu\nu}[\xi|s]$:
\begin{equation}
\epsilon^{\mu\nu\rho\sigma} {\cal D}_\rho(s) G_{\mu\nu}[\xi|s] = 0
\label{loopbianchi}
\end{equation}
implies by (\ref{Gausslawldc}) that the loop covariant derivative of
$\kappa[\xi|s]$ in (\ref{dtracelk}) vanishes:
\begin{equation}
{\cal D}_\mu(s) \kappa[\xi|s] = 0.
\label{curldkappa}
\end{equation}
Hence, we can use (\ref{ELeq1}) to eliminate the Lagrange multiplier
$L_{\mu\nu}[\xi|s]$ from (\ref{ELeq2}) obtaining:
\begin{equation}
m \frac{d^2Y_\mu(\tau)}{d\tau^2} = \frac{2 {\tilde g}}{{\bar N}} \int
   \delta \xi ds \epsilon_{\mu\nu\rho\sigma} \Tr\{\kappa[\xi|s] F^\nu[\xi|s]\}
   \dot{\xi}^\rho(s) \frac{d Y^\sigma(\tau)}{d\tau} \dot{\xi}(s)^{-2}
   \delta^4(\xi(s)-Y(\tau)),
\label{produalwong}
\end{equation}
which is reminiscent of the Lorentz equation.  Indeed, if we substitute for
$F_\mu[\xi|s]$ the expression in (\ref{Fmuxisx}), we obtain:
\begin{equation}
m \frac{d^2 Y_\mu(\tau)}{d\tau^2} = - {\tilde g} \Tr[K(\tau)
\hodge F_{\mu\nu}(Y(\tau))]
   \frac{d Y^\nu(\tau)}{d\tau},
\label{dualwong}
\end{equation}
which is exactly the dual of the Wong equation \cite{Wong} for the classical
limit of a Yang-Mills source.

The equations (\ref{Gausslawldc}), (\ref{Polyakov}), and (\ref{dualwong})
are the same as those derived previously \cite{Chanstsou2} with the global
version (\ref{Gausslawl}) of the topological constraint defining the
$\zeta$-charge, affording thus a good internal consistency check for our
method.

\section{Quantum Mechanics of $\zeta$-Charge}

Having developed the necessary techniques and understanding by working
through the quantum abelian theory as well as the classical nonabelian case,
we now turn to our target problem of a Dirac particle carrying a nonabelian
$\zeta$-charge.  The action remains (\ref{freeaction}) with ${\cal A}^0_F$,
in view of patching, written in terms of loop variables as (\ref{freefieldal}),
and ${\cal A}^0_M$ given as in (\ref{freeqparta}).  Equations of motion
are to be derived by extremizing this action with respect to the variables
$F_\mu[\xi|s]$ and $\psi(x)$ subject to the topological constraint that
the particle carries a $\zeta$-charge.

For the explicit form of the constraint, we seek to replace the classical
$\zeta$-current in (\ref{Gausslawldc}) by the quantum equivalent as we did
for the abelian theory in (\ref{gausslawq}).  What, however, is the classical
$\zeta$-current?  In view of the similarity in appearance of (\ref{Gausslawdc})
to the Wong equation \cite{Wong} for the classical Yang-Mills source, one might
be tempted to interpret its right-hand side as the $\zeta$-current, but this
would be incorrect.  Like the `source' current in the Wong equation, the
right-hand side of (\ref{Gausslawdc}) can readily be seen to be covariantly
conserved, namely having vanishing covariant divergence.  Now the (`colour')
`source' current in the ordinary Yang-Mills theory is covariantly conserved
because the gauge field itself carries a (`colour') source charge and it is
the total (`colour') source charge which is to be conserved.  However, in the
present case, the field has no $\zeta$-charge which is carried only by the
particle at $Y(\tau)$.  Hence, the $\zeta$-current carried by $Y(\tau)$ should
by itself be already a conserved quantity, or in other words it should have
a vanishing ordinary (as opposed to covariant) divergence.  For this
reason, we see that the right-hand side of (\ref{Gausslawdc}) could not
be taken as the $\zeta$-current carried by the $\zeta$-charge at $Y(\tau)$.

Nevertheless, from the right-hand side of (\ref{Gausslawdc}), we can construct
a properly conserved $\zeta$-current simply by replacing $K(\tau)$ there
by another quantity ${\cal K}(\tau)$, thus:
\begin{equation}
{\tilde j}_\mu (x) = - {\tilde g} \int d\tau {\cal K}(\tau) \frac{dY^\mu(\tau)}
   {d\tau} \delta^4(x-Y(\tau))
\label{zetacurrent}
\end{equation}
where:
\begin{equation}
{\cal K}(\tau) = \left. \Omega_\xi(s,0) \kappa[\xi|s] \Omega^{-1}(s,0)
   \right |_{\xi(s) = Y(\tau)},
\label{curlyk}
\end{equation}
with:
\begin{equation}
\Omega(s,0) = \omega(\xi(s_+)) \Phi_\xi(s_+,0).
\label{Omega}
\end{equation}
To explain what is meant, we recall first that in the definition (\ref{Ktau})
of $K(\tau)$, the factor $\Phi_\xi(s_-,0)$ is a parallel transport from the
loop space reference point $P_0$ to the point $\xi(s_-)$ so that the
`phase' or orientation of $K(\tau)$ in internal symmetry space is measured
in the local frame at $\xi(s_-)$.  Similarly, the factor $\Omega_\xi(s,0)$
in (\ref{curlyk}) and (\ref{Omega}) is also a parallel transport, but now
from the reference $P_0$ to the point $\xi(s_+)$, where $s_+$ means that
the value of $\xi(s)$ has now to be taken as the limit from above, namely
$\xi(s_+) = \xi(s + \epsilon), \epsilon \rightarrow 0+$.  The symbol
$\omega(x)$ introduced in (\ref{Omega}) represents just a local, i.e.
$x$-dependent, rotation matrix to allow for the possibility that the `phase'
of the $\zeta$-current may be measured in a local frame different from that
in which the field is measured; it will be of relevance later when considering
`phase' rotations of $\zeta$-charges but does not enter in the present
question of current conservation.  What makes the difference in this question
is the replacement of the argument $s_-$ in $\Phi_\xi(s_-,0)$ of (\ref{Ktau})
by $s_+$ in $\Phi_\xi(s_+,0)$ of (\ref{Omega}).  Pictorially, this is as
indicated in Figure \ref{diffloop}, where it is clear that whereas the factor
\begin{figure}
\vspace{7cm}
\caption{Loop derivative of $\Omega_\xi(s,0)$ and $\Phi[\xi]$}
\label{diffloop}
\end{figure}
$\Phi_\xi(s_-,0)$ in (\ref{Ktau}) depends on $\xi(s')$ only for $s'$ up to
$s_-$ and is therefore unaffected by the loop derivative $\delta/\delta
\xi^\mu(s)$ at $s$, this is no longer true for $\Omega_\xi(s,0)$ or
$\Phi_\xi(s_+,0)$.  Indeed, the logarithmic derivative with respect to
$\xi(s)$ of $\Omega_\xi(s,0)$ on the left-hand side of (\ref{difomegaxi}) below
may be represented pictorially by the solid curve in Figure \ref{diffloop},
which is in fact the same as $ -ig F_\mu[\xi|s]$ in (\ref{Fmuxis}), the
logarithmic derivative of $\Phi[\xi]$, since the latter is represented by
the whole curve in Figure \ref{diffloop} including the dotted segments, but
these by our ordering convention will cancel in any case.  Hence, we have:
\begin{equation}
\Omega^{-1}_\xi(s,0) \frac{\delta}{\delta \xi^\mu(s)} \Omega_\xi(s,0)
   = -ig F_\mu[\xi|s].
\label{difomegaxi}
\end{equation}
That being the case, it follows then from (\ref{curldkappa}) that the
$\zeta$-current defined in (\ref{zetacurrent}) has indeed a vanishing
ordinary divergence as we wanted.

In going over then from the classical theory into the quantum theory, one
follows the example worked out in the abelian case and replaces the classical
$\zeta$-current (\ref{zetacurrent}) by its quantum analogue, thus:
\begin{equation}
{\tilde j}_\mu (x) = - {\tilde g} [{\bar \psi}(x) \gamma_\mu T^i \psi(x)]
   \tau_i,
\label{zetacurrentq}
\end{equation}
where for $\psi(x)$ belonging to a representation of the gauge group and $T^i$
being a matrix representing the generators $\tau_i$ in this representation, the
current is an element of the gauge algebra as it should.  As a result, we
have the constraint:
\begin{equation}
G_{\mu\nu}[\xi|s] = 4\pi {\tilde g} \epsilon_{\mu\nu\rho\sigma}
   \dot{\xi}^\rho(s) [{\bar \psi}(\xi(s)) \gamma^\sigma T^i \psi(\xi(s))]
   \Omega_\xi^{-1}(s,0) \tau_i \Omega_\xi(s,0) ,
\label{Gausslawldq}
\end{equation}
which is to be imposed for deriving the quantum equations of motion.

Incorporating the constraint into the action, one has again (\ref{Fullact})
as before, but with $J_{\mu\nu}[\xi|s]$ now given by $-1/4\pi$ times the
right-hand side of (\ref{Gausslawldq}).  Extremizing then (\ref{Fullact})
with respect to $F_\mu[\xi|s]$, we obtain (\ref{ELeq1}) as before, but with
respect to $\psi(x)$:
\begin{equation}
(i \partial_\mu \gamma^\mu - m) \psi(x) = - {\tilde g} {\tilde A}_\mu(x)
   \gamma^\mu \psi(x),
\label{dualdirac}
\end{equation}
where:
\begin{equation}
{\tilde A}_\mu(x) = 4\pi \int \delta \xi ds \epsilon_{\mu\nu\rho\sigma}
   \Omega_\xi(s,0) L^{\rho\sigma}[\xi|s] \Omega_\xi^{-1}(s,0) \dot{\xi}^\nu(s)
   \delta^4(x - \xi(s)).
\label{Atilde}
\end{equation}
Together then with (\ref{Gausslawldq}) and (\ref{ELeq1}), (\ref{dualdirac})
completes the set of equations governing the motion of a Dirac particle with
$\zeta$-charge moving in the Yang-Mills field.

One sees that the equation (\ref{dualdirac}) is formally the same as the
ordinary Yang-Mills equation for a `colour' source moving in a gauge field.
As in the abelian theory, a new local quantity ${\tilde A}_\mu(x)$
has emerged coupling to the $\zeta$-charge's wave function in just the
same manner as the usual potential $A_\mu(x)$ to the wave function of a
source charge.  As in the abelian theory also, this ${\tilde A}_\mu(x)$
appears via the Lagrange multiplier to the topological constraint defining
the $\zeta$-charge, although it is no longer just proportional to the
Lagrange multiplier but related to it in a more complicated fashion through
(\ref{Atilde}).  We shall still call ${\tilde A}_\mu(x)$ the {\it dual
potential} although it should be stressed that it is not the potential
in the usual sense of the dual tensor $\hodge F_{\mu\nu}(x)$; namely it
need not satisfy:
\begin{equation}
\hodge F_{\mu\nu}(x) = \partial_\nu {\tilde A}_\mu(x) - \partial_\mu
   {\tilde A}_\nu(x) + ig [{\tilde A}_\mu(x), {\tilde A}_\nu(x)],
\label{wrongdual}
\end{equation}
which, as we know, does not usually have solutions.  Indeed, as we
shall see, ${\tilde A}_\mu(x)$ has a much more intricate relationship with the
field but still performs, though in a rather unusual fashion, as a parallel
`phase' transport for the $\zeta$-charge.

\section{Chiral Doubling of Symmetry}

In the abelian theory, we have seen that we have a `chiral' doubling of
the gauge symmetry from the original $U(1)$ to $U(1) \times U(1)$, where
the second $U(1)$ (henceforth referred to as ${\tilde U}$) carries a parity
opposite to that of the first.  This
is in a sense expected since the abelian theory is dual symmetric, and
what applies to sources must also apply to monopoles, and since there is
a $U$-invariance associated with the phase of sources, so there must also
be a ${\tilde U}$-invariance associated with the phase of monopoles.

For the nonabelian theory, one has no dual symmetry between source charges and
$\zeta$-charges.  Hence, the fact that one has a $U$-invariance (i.e. the
original Yang-Mills gauge invariance) associated with the `phase' of
source charges does not necessarily imply the existence of a corresponding
${\tilde U}$-invariance associated with the `phase' of $\zeta$-charges.
Nevertheless, given the similarity in formulation above between the
abelian and nonabelian theories, it might be conjectured that even the
nonabelian theory has a ${\tilde U}$-invariance.  This turns out to be
indeed the case although the ${\tilde U}$-invariance, as we shall see,
is realized here in a rather novel and more intricate fashion.

By a ${\tilde U}$ transformation we mean a local rotation in `phase' of
the wave function $\psi(x)$ of particles carrying $\zeta$-charges, thus:
\begin{equation}
\psi(x) \longrightarrow (1 + i {\tilde g} {\tilde \Lambda}(x)) \psi(x)
\label{Utildeonpsi}
\end{equation}
for an arbitrary (infinitesimal) algebra-valued function ${\tilde \Lambda}(x)$.
This is different from the ordinary Yang-Mills $U$-transformation, parametrized
say by $\Lambda(x)$, under which the wave function $\psi(x)$ of the
$\zeta$-charge, being `colour' neutral, is invariant.  By construction,
the field variables $F_\mu[\xi|s]$ are invariant under a simultaneous
$U$- and ${\tilde U}$-transformation (apart from a harmless $x$-independent
$U$-rotation at the reference point $P_0$ which we have agreed above in
Section 3 to ignore), while the rotation matrices $\omega(x)$ and
$\Omega_\xi(s,0)$ introduced in the section above will transform as:
\begin{equation}
\omega(x) \longrightarrow (1 + i{\tilde g} {\tilde \Lambda}(x)) \omega(x)
   (1 - ig \Lambda(x)),
\label{omegatransf}
\end{equation}
and:
\begin{equation}
\Omega_\xi(s,0) \longrightarrow [1 + i{\tilde g}{\tilde \Lambda}(\xi(s_+))]
   \Omega_\xi(s,0).
\label{Omegatransf}
\end{equation}
There remains then, of the quantities that we have introduced, only the
Lagrange multiplier $L_{\mu\nu}[\xi|s]$, and associated with it the
`dual potential' ${\tilde A}_\mu(x)$, for which the transformation properties
have yet to be specified.

Under pure $U$-transformations, $L_{\mu\nu}[\xi|s]$ may be taken to be
invariant and so also, by (\ref{Omegatransf}), may ${\tilde A}_\mu(x)$.
Under ${\tilde U}$-transformations, on the other hand, we have the freedom
to choose such transformation laws for $L_{\mu\nu}[\xi|s]$ so as to leave
the action ({\ref{Fullact}) invariant.  From the actual structure of
(\ref{Fullact}), we see that if ${\tilde A}_\mu(x)$ transforms under
${\tilde U}$ analogously to an ordinary potential under $U$, namely as:
\begin{equation}
{\tilde \Delta}{\tilde A}_\mu(x) = \partial_\mu {\tilde \Lambda}(x)
   +i {\tilde g} [{\tilde \Lambda}(x), {\tilde A}_\mu(x)],
\label{Atildetransf}
\end{equation}
then, the increment due to ${\tilde \Delta} {\tilde A}_\mu(x)$ in the last
term of (\ref{Fullact}):
\begin{equation}
4\pi \int \delta \xi ds \Tr \{L^{\mu\nu}[\xi|s] J_{\mu\nu}[\xi|s] \}
   = {\tilde g} \int d^4x {\bar \psi}(x) {\tilde A}_\mu(x) \gamma^\mu \psi(x)
\label{interaction}
\end{equation}
will exactly cancel in a familiar fashion with the corresponding change in
the particle free action ${\cal A}^0_M$ of (\ref{freeqparta}) due to the
transformation (\ref{Utildeonpsi}) of $\psi(x)$.  Hence, if we choose
to have $L_{\mu\nu}[\xi|s]$ transforming in such a way as to give
${\tilde A}_\mu(x)$ in (\ref{Atilde}) the transformation (\ref{Atildetransf})
and at the same time leave the term proportional to $\Tr \{L^{\mu\nu}[\xi|s]
G_{\mu\nu}[\xi|s]\}$ in (\ref{Fullact}) invariant, then the whole theory
will be invariant under ${\tilde U}$.

The following transformation for $L_{\mu\nu}[\xi|s]$ satisfies the above
requirements:
\begin{equation}
{\tilde \Delta} L_{\mu\nu}[\xi|s] = \epsilon_{\mu\nu\rho\sigma}
   {\cal D}^\rho(s) {\tilde \Lambda}^\sigma[\xi|s],
\label{Lmunutransf}
\end{equation}
as can be seen as follows.  On substitution into (\ref{Fullact}), one finds
after integration by parts that:
\begin{equation}
\int \delta \xi ds \Tr \{{\tilde \Delta}L^{\mu\nu}[\xi|s] G_{\mu\nu}[\xi|s] \}
   = - \int \delta \xi ds \Tr \{\epsilon^{\mu\nu\rho\sigma}
   {\tilde \Lambda}_\sigma[\xi|s] {\cal D}_\rho(s) G_{\mu\nu}[\xi|s]\},
\label{Utonfielda}
\end{equation}
which vanishes by the Bianchi identity (\ref{loopbianchi}) of
$G_{\mu\nu}[\xi|s]$.  Further, on substituting (\ref{Lmunutransf}) into
(\ref{Atilde}) and using (\ref{Omegatransf}), we have:
\begin{equation}
{\tilde \Delta}{\tilde A}_\mu(x) = i{\tilde g} [{\tilde \Lambda}(x),
   {\tilde A}_\mu(x)] + {\tilde \Delta}_L {\tilde A}_\mu(x),
\label{Atildeprotr}
\end{equation}
with:
\begin{equation}
{\tilde \Delta}_L {\tilde A}_\mu(x) = - 8\pi \int \delta \xi ds
   \Omega_\xi(s,0) {\cal D}_\mu {\tilde \Lambda}_\nu[\xi|s]
\Omega_\xi^{-1}(s,0)
   \dot{\xi}^\nu(s) \delta^4(\xi(s) - x).
\label{Deltatlat}
\end{equation}
Then using the relation (\ref{difomegaxi}), we have:
\begin{equation}
\Omega_\xi(s,0) \{{\cal D}_\mu(s) {\tilde \Lambda}_\nu[\xi|s]\}
   \Omega^{-1}_\xi(s,0) = \frac{\delta}{\delta \xi^\mu(s)} \{\Omega_\xi(s,0)
   {\tilde \Lambda}_\nu[\xi|s] \Omega_\xi^{-1}(s,0)\},
\label{dandcurlyd}
\end{equation}
which on substitution into (\ref{Atildeprotr}) gives (\ref{Atildetransf})
as desired, so long as one defines:
\begin{equation}
{\tilde \Lambda}(x) = - 8 \pi \int \delta \xi ds \Omega_\xi(s,0)
   {\tilde \Lambda}_\nu[\xi|s] \Omega^{-1}_\xi(s,0) \dot{\xi}^\nu(s)
   \delta^4(\xi(s) - x).
\label{Lambdatilde}
\end{equation}

We have thus shown that the action (\ref{Fullact}) has indeed also a
${\tilde U}$-invariance, or that the symmetry is doubled in the nonabelian
theory as in the abelian case, in spite of the fact that the former has
no dual symmetry.  The physical origin of this ${\tilde U}$-invariance
can be traced back to the fact that the $\zeta$-charge was initially defined
only as a homotopy class of closed curves in the gauge group $G$ and as such
has no orientation
or `phase' in internal symmetry space.  Thus, for example, for the ${\bf
su}(2)$
theory, the $\zeta$-charge is labelled only by a sign.  In formulating the
dynamics, however, as was done in Sections 4 and 5, the $\zeta$-charge was
assigned a `phase', either in the classical case by the quantity $K(\tau)$
or in the quantum case by the representation of $G$ to which the wave function
$\psi(x)$ belongs, which `phase', however, the $\zeta$-charge did not
originally possess.  It follows therefore that this `phase' ought to be
unphysical, and can be redefined arbitrarily at any space-time point
without altering the physics.  In other words, we expect the theory to
be invariant under local `phase' rotations of the $\zeta$-charge, which
is just the statement of the above ${\tilde U}$-invariance.

One notices, however, that the ${\tilde U}$ transformations above are
parametrized by a vector-function ${\tilde \Lambda}_\mu[\xi|s]$ carrying an
index $\mu$, not by a scalar function as usual gauge transformations are.
The reason for this is that the `phase' was assigned to the $\zeta$-charge
through (\ref{Gausslawldc}) or (\ref{Gausslawldq}) which depends on 3
space-time
indices, namely the indices $\mu, \nu$ of the loop curvature
$G_{\mu\nu}[\xi|s]$
and the index $\rho$ for the tangent to the loop $\dot{\xi}^\rho(s)$, which
together were needed to specify the orientation in space-time of the elemental
3-volume enclosed by the little surface in Figure \ref{smallloop}.  In
4-dimensions, this is equivalent to saying that the `phase' assigned to
the $\zeta$-charge depends on a space-time direction which may be taken to
be the normal to the elemental 3-volume in Figure \ref{smallloop} and
represented by the index $\sigma$ in (\ref{Gausslawldc}) and
(\ref{Gausslawldq}).  Given that the physics should be independent of
this `phase', one should have therefore the freedom to redefine the
`phase' arbitrarily for each direction also; hence the extra index on
the gauge parameter ${\tilde \Lambda}_\mu[\xi|s]$.

This ${\tilde U}$ invariance being also a `local' symmetry, there ought to
be `gauge potential' or `connection' to specify parallel `phase' transport.
This is supplied by the Lagrange multiplier $L_{\mu\nu}[\xi|s]$, in close
analogy to the abelian case of Section 2 where it was again the Lagrange
multiplier $\lambda_\mu(x)$ which supplied the `connection' for parallel phase
transport of monopoles.  However, since the `phase' now depends on a direction,
the `connection' has to carry an extra index so as to specify what is meant
by parallel `phases' not only at neighbouring points but also in
neighbouring directions.  Such 2-indexed tensor potentials have already been
considered before in a different context, originally in string and
supersymmetry
theories \cite{Hayashi,Kalbramond,Cremmerscherk,Nambu}.  Indeed, the
transformation law (\ref{Lmunutransf}) is just the loop space version of the
transformation law for nonabelian tensor potentials first suggested by Freedman
and Townsend in 1981 \cite{Freetown,Chanftsouw}.  The space-time local `dual
potential' ${\tilde A}_\mu(x)$ and its corresponding `gauge parameter'
${\tilde \Lambda}(x)$ defined above in (\ref{Atilde}) and (\ref{Lambdatilde})
are obtained essentially by taking the directional average of the
corresponding quantities with an extra index, namely $L_{\mu\nu}[\xi|s]$
and ${\tilde \Lambda}_\mu[\xi|s]$ respectively.

\section{Remarks}

As said in the introduction, the ultimate object of our program is hopefully
to learn sufficiently about the dynamical properties of $\zeta$-charges to
eventually decide whether they might exist in Nature.  Obviously, we are
still quite far from reaching that goal.  However, the derivation of the
the quantum equations of motion should be a significant step forward since
almost all applications of non-abelian Yang-Mills theories have so far been in
quantum physics.  Knowing now the quantum equations of what one might call the
`dual Yang-Mills theory', we may then hopefully begin to explore their
consequences.

In view of possible applications, it may be useful to find out what the
present equations in unfamiliar loop space form may correspond to in usual
space-time notation by substituting into them the expressions such as
(\ref{Fmuxisx}) in Section 3.  In each case, whether classical or quantum,
two equations are known already in space-time local notation, namely
(\ref{dualwong}) and (\ref{Gausslawdc}) for the classical, but
(\ref{dualdirac}) and a similar equation to (\ref{Gausslawdc}) with a quantum
$\zeta$-current for the quantum case.  There remains then only the equation
(\ref{ELeq1}), which is common to both cases.  On substituting (\ref{Atilde})
into (\ref{ELeq1}) and using (\ref{difomegaxi}), we obtain:
\begin{equation}
F_{\mu\nu}(x) = \epsilon_{\mu\nu\rho\sigma} \omega^{-1}(x) \{\partial^\sigma
   {\tilde A}^\rho(x)\} \omega(x).
\label{ELeq1x}
\end{equation}
This last equation is a little peculiar in that it does not seem to be, but
actually is, covariant under a ${\tilde U}$-transformation as it ought to
be, having been derived from a loop equation which was known to be covariant.
The reason for this apparent paradox is that a ${\tilde U}$-transformation,
as explained in the last section, depends not only on a point in space-time
but also on a direction.  Hence, the derivative of ${\tilde A}_\mu(x)$, which
was obtained as a directional average of the tensor potential
$L_{\mu\nu}[\xi|s]$, does not transform as usual potentials do under an
ordinary local transformation, but rather as follows:
\begin{equation}
\partial_\sigma {\tilde A}_\rho \longrightarrow
   (1 + i{\tilde g} {\tilde \Lambda}(x)) [\partial_\sigma {\tilde A}_\rho(x)
   + \partial_\sigma \partial_\rho {\tilde \Lambda}(x)]
   (1 - i{\tilde g} {\tilde \Lambda}(x)),
\label{difattransf}
\end{equation}
leaving then the equation (\ref{ELeq1x}) covariant under ${\tilde U}$ as
it should.  Although when compared with the original loop space versions,
these equations in space-time look simpler, we are unsure of their
significance in that they involve via the covariant derivative $D_\mu$ the
ordinary Yang-Mills potential $A_\mu(x)$ which is undefined at the position
of the $\zeta$-charge.  For this reason, we have still preferred up to the
present to work with the equations in loop space.

We note that although one has derived explicitly the equations of motion only
for one single particle carrying a $\zeta$-charge moving in the Yang-Mills
field, the derivation can immediately be extended to any finite number of
similar particles, whether in the classical or the quantum theory.  The
resulting equations may then be used to study the interaction between
$\zeta$-charges.  Or else, in the quantum case, one may imagine the wave
function $\psi(x)$ to be a quantized field $\psi(x)$, and use the equations
as launch-pads for exploring say the scattering of one $\zeta$-charge from
another.  Preliminary investigations suggest that some properties of
interacting $\zeta$-charges remain superficially similar to ordinary
source charges.  One immediate difference, however, is that whereas the
source charge couples to the gauge field with the same coupling as the
gauge field self-coupling $g$, the $\zeta$-charge couples to the gauge field
via a coupling ${\tilde g}$ which is related to $g$ by the (extended) Dirac
quantization condition, which for $SU(N)$ theory reads as \cite{Chantsou}:
\begin{equation}
g {\tilde g} = n/2N.
\label{Diracquant}
\end{equation}
This difference in coupling is probably a first signature to look for when
trying to ascertain whether a particle can be a $\zeta$-charge.

On thing we cannot do as yet, however, is to consider interactions between
source charges and $\zeta$-charges.  The difficulty for us is that in order to
specify the coupling of source charges one requires, it seems, the gauge
potential $A_\mu(x)$ so that one cannot avoid then the complications of
patching or of the Dirac string as one has done in the problem above by
changing over into gauge-idependent loop variables.  This is a serious
limitation to our present program which may not be easy to overcome.

\subsection*{Acknowledgement}

One of us (JF) acknowledges the support of the Mihran and Azniv Essefian
Foundation (London), the Soudavar Foundation (Oxford) and the Calouste
Gulbenkian Foundation (Lisbon), while another (TST) thanks the Windgate
Foundation for partial support during the latter part of this work.

\end{document}